\pdfoutput=1
\documentclass[twocolumn,prl,superscriptaddress,longbibliography]{revtex4-2}
\usepackage[colorlinks=true, citecolor=blue, urlcolor=blue, linkcolor=red]{hyperref}
\renewcommand{\section}[1]{{\par\it #1.---}\ignorespaces}
\usepackage{amsmath,amssymb,tikz,scalerel}
\hypersetup{
        colorlinks=true,
        linkcolor=red,
        citecolor=blue,
        urlcolor=blue}
\usetikzlibrary{svg.path}
\definecolor{orcidlogocol}{HTML}{A6CE39}
\tikzset{
	orcidlogo/.pic={
		\fill[orcidlogocol] svg{M256,128c0,70.7-57.3,128-128,128C57.3,256,0,198.7,0,128C0,57.3,57.3,0,128,0C198.7,0,256,57.3,256,128z};
		\fill[white] svg{M86.3,186.2H70.9V79.1h15.4v48.4V186.2z}
		svg{M108.9,79.1h41.6c39.6,0,57,28.3,57,53.6c0,27.5-21.5,53.6-56.8,53.6h-41.8V79.1z M124.3,172.4h24.5c34.9,0,42.9-26.5,42.9-39.7c0-21.5-13.7-39.7-43.7-39.7h-23.7V172.4z}
		svg{M88.7,56.8c0,5.5-4.5,10.1-10.1,10.1c-5.6,0-10.1-4.6-10.1-10.1c0-5.6,4.5-10.1,10.1-10.1C84.2,46.7,88.7,51.3,88.7,56.8z};}}
\newcommand\orcid[1]{\href{https://orcid.org/#1}{\mbox{\scalerel*{\begin{tikzpicture}[yscale=-1,transform shape]\pic{orcidlogo};\end{tikzpicture}}{|}}}}

\begin{document}
\title{Semimetals without correspondence between the topological charge of nodal line/surface and Fermi arc}
\author{Xue-Min Yang\orcid{0000-0002-6937-8402}}
\affiliation{School of Science, Chongqing University of Posts and Telecommunications, Chongqing 400065, China}
\affiliation{Institute for Advanced Sciences, Chongqing University of Posts and Telecommunications, Chongqing 400065, China}
\author{Jiang-Shan Chen}
\affiliation{School of Science, Chongqing University of Posts and Telecommunications, Chongqing 400065, China}
\affiliation{Institute for Advanced Sciences, Chongqing University of Posts and Telecommunications, Chongqing 400065, China}
\author{Ying Chen}
\affiliation{School of Science, Chongqing University of Posts and Telecommunications, Chongqing 400065, China}
\author{Yu-Hang Qin}
\affiliation{School of Science, Chongqing University of Posts and Telecommunications, Chongqing 400065, China}
\author{Hong Wu\orcid{0000-0003-3276-7823}}\email{Corresponding author: wuh@cqupt.edu.cn}
\affiliation{School of Science, Chongqing University of Posts and Telecommunications, Chongqing 400065, China}
\affiliation{Institute for Advanced Sciences, Chongqing University of Posts and Telecommunications, Chongqing 400065, China}

\begin{abstract}
It is generally believed that there is a correspondence between the topological charge of nodal points or lines and the presence of Fermi arcs. Using a $\mathcal{P}\mathcal{T}$-invariant system as an example, we demonstrate that this general belief is no longer valid. When two charged nodal lines or surfaces touch, the topological charge {dissipates without gap opening}, yet the surfaces or hinge Fermi arcs can remain preserved. It is found that both static and Floquet semimetals can exhibit Fermi arcs, even when the nodal lines or surfaces do not carry a $Z_2$ charge from second Stiefel-Whitney class. 
\end{abstract}
\maketitle
\section{Introduction}
 Due to potential applications and new paradigm in condensed matter physics brought by topological phases, these quantum states have garnered widespread attention over the past decades \cite{RevModPhys.83.1057,RevModPhys.91.015006,RevModPhys.82.3045,RevModPhys.91.015005,RevModPhys.93.025002,RevModPhys.90.015001}. The core feature of these phases is the principle of bulk-boundary correspondence. The existence of boundary states is dictated by the bulk topological {invariants,} which are defined by considering various internal and space group symmetries \cite{RevModPhys.88.035005,PhysRevX.9.021013,PhysRevX.9.011012,PhysRevLett.119.246401}. Based on this classification rule, topological insulators, semimetals, and superconductors featuring symmetry-protected boundary states have been discovered \cite{PhysRevLett.122.014302,PhysRevLett.112.216803,PhysRevB.76.045302,PhysRevLett.109.227003,Benalcazar_2017,PhysRevLett.118.076803,Schindler_2018,PhysRevB.83.205101,PhysRevLett.125.146401,PhysRevLett.125.266804,PhysRevLett.122.236401,PhysRevX.5.031013,PhysRevLett.108.140405,PhysRevLett.113.027603,PhysRevB.84.235126}. 

{A topological semimetal is a quantum phase characterized by symmetry-protected band crossings (such as nodal points \cite{PhysRevB.92.081201,PhysRevLett.121.106404,PhysRevLett.132.016603}, nodal lines \cite{PhysRevB.96.041103,PhysRevB.96.201305,PhysRevLett.118.045701}, or nodal surfaces \cite{PhysRevLett.132.186601,PhysRevB.111.045302}) near the Fermi level.} Recently, topological semimetals have been widely studied in systems with space-time ($\mathcal{P}\mathcal{T}$) inversion symmetry due to {their} importance for a broad class of interesting systems, including not only solid-state quantum materials \cite{PhysRevX.9.021013,PhysRevLett.123.256402,wang2024mirrorrealcherninsulator} but also photonic, cold-atom, classical acoustic, and circuit systems \cite{RevModPhys.91.015006,PhysRevLett.114.114301,Lee_2020,PhysRevA.109.053314}. First, {such systems} can exhibit multi-gap topologies with band nodes that carry non-Abelian charges, characterized by the braiding of these nodes in momentum space \cite{Wu_2019,Guo_2021,sun2023nonabeliantopologicalphasesquotient,Jiang_2021,Bouhon_2020,pengbo}. {Additionally,} the Dirac point in {these systems} is unstable and will generically transform into a nodal line when the $\mathcal{P}\mathcal{T}$-invariant relevant perturbations are turned on \cite{PhysRevB.92.081201,PhysRevLett.125.126403}. Because the nodal lines originate from Dirac points, they both host the same $Z_2$ charge, defined by the second Stiefel-Whitney class on a sphere enclosing each nodal point or loop \cite{PhysRevLett.125.126403,PhysRevLett.121.106403,PhysRevX.8.031069}. 
Then, three-dimensional (3D) semimetals can be sliced into a family of 2D $k_z$-dependent normal and topological insulators.  The topological charge of each nodal point or line {equals} the difference in topological invariants between its separated phases \cite{PhysRevLett.125.266804}. The topological boundary modes of all $k_z$-dependent 2D topological insulators contribute to Fermi arcs that connect nodal points or lines with opposite charges \cite{PhysRevLett.125.126403,PhysRevLett.128.026405}. {These studies suggest} a correspondence between the topological charge of the nodal points/lines and the Fermi arcs.

In this work, we investigate the $\mathcal{P}\mathcal{T}$-invariant semimetals and their Floquet engineering. It is intriguing to find that the correspondence between the topological charge of nodal points/lines and Fermi arcs can be broken. {These} semimetals can exhibit hinge Fermi arcs even if the nodal lines do not carry a $Z_2$
charge. Unlike the conventional case, where nodal lines separate 2D normal and topological insulators, the nodal lines in our system separate topological insulators with the same bulk invariants. {Through} Floquet engineering, we discover exotic nodal surface semimetal{s} with surface Fermi arc{s}. These nodal surfaces still {carry} no $Z_2$ charge. 

\section{Stiefel-Whitney classes and topological phases in $\mathcal{P}\mathcal{T}$-invariant systems} The fundamental symmetry in our system is space-time ($\mathcal{P}\mathcal{T}$) inversion symmetry. The symmetry operator $\mathcal{P}\mathcal{T}$ can always be represented as the complex conjugation operator $\mathcal{K}$ with a suitable choice of {basis, leading to the relation} $\mathcal{P}\mathcal{T} \mathcal{H}(\mathbf{k}) (\mathcal{P}\mathcal{T})^{-1}=\mathcal{H}^{*}(\mathbf{k})=\mathcal{H}(\mathbf{k})$. {Such systems are} topologically classified by Stiefel-Whitney numbers \cite{PhysRevLett.121.106403}. Over a closed 1D manifold chosen within 3D Brillouin zone, the $Z_2$ topological invariant {measuring} the orientability of real
quantum states is first Stiefel-Whitney number $w_1$:
 \begin{equation}
w_1=\frac{1}{\pi}\oint_C d\mathbf{k}.\text{Tr}[A(\mathbf{k})],
\end{equation}
where $A_{mn}=\langle u_{m\mathbf{k}} \lvert i\triangledown_{\mathbf{k}} \lvert u_{n\mathbf{k}} \rangle$ is the real Berry connection calculated in a complex smooth gauge, and $C$ is a closed curve in the Brillouin zone \cite{PhysRevLett.121.106403}. {Here} $\lvert u_{n\mathbf{k}} \rangle$ is the eigenstate of $\mathcal{H}(\mathbf{k})$. Over a closed 2D manifold within {the} 3D Brillouin zone, {the system} is topologically classified by the second Stiefel-Whitney number $w_2$, {which} can be computed by the Wilson-loop method. By choosing a sphere enclosing the band touching line or surface, the Wilson loop operator is calculated along $\phi$ at each $\theta$ over the sphere, where $(k_x-k'_x,k_y-k'_y,k_z-k'_z)$=$\lvert r \lvert(\sin \theta\cos \phi, \sin\theta \sin\phi, \cos\theta)$. Then 
\begin{equation}
W(\theta)=P e^{-i\int_{C_{\theta}}A(\phi)d\phi}=\Pi_{i=0}^{N_{\phi}-1}F_{i,i+1}(\theta),
\end{equation}
where $P$ denotes path ordering, $C_{\theta}$ is the contour at fixed $\theta$, and $A$ is the Berry connection. $F_{i,i+1}^{mn}(\theta)=\langle   u_m  (\phi_{i},\theta)  \lvert    u_n(\phi_{i+1},\theta)    \rangle$ with $F_{N_{\phi}-1,N_{\phi}}=F_{N_{\phi}-1,0}$. The topology is encoded in the phase factors $\varphi_m(\theta)$ of $\lambda_m(\theta)$, which are the eigenvalues for $W(\theta)$. We can obtain Wilson-loop spectrum in $\lambda-\theta$ plane through calculations. Finally, $w_2$ can be read off from $\lambda_m(\theta)$ as the parity of
the number of crossing points at $\varphi_m(\theta)$=$\pi$ \cite{PhysRevLett.121.106403}. When $w_2=1$ ($0$), the nodal points or lines (don't) carry a $Z_2$ charge.

\begin{figure}[tbp]
\centering
\includegraphics[width=1\columnwidth]{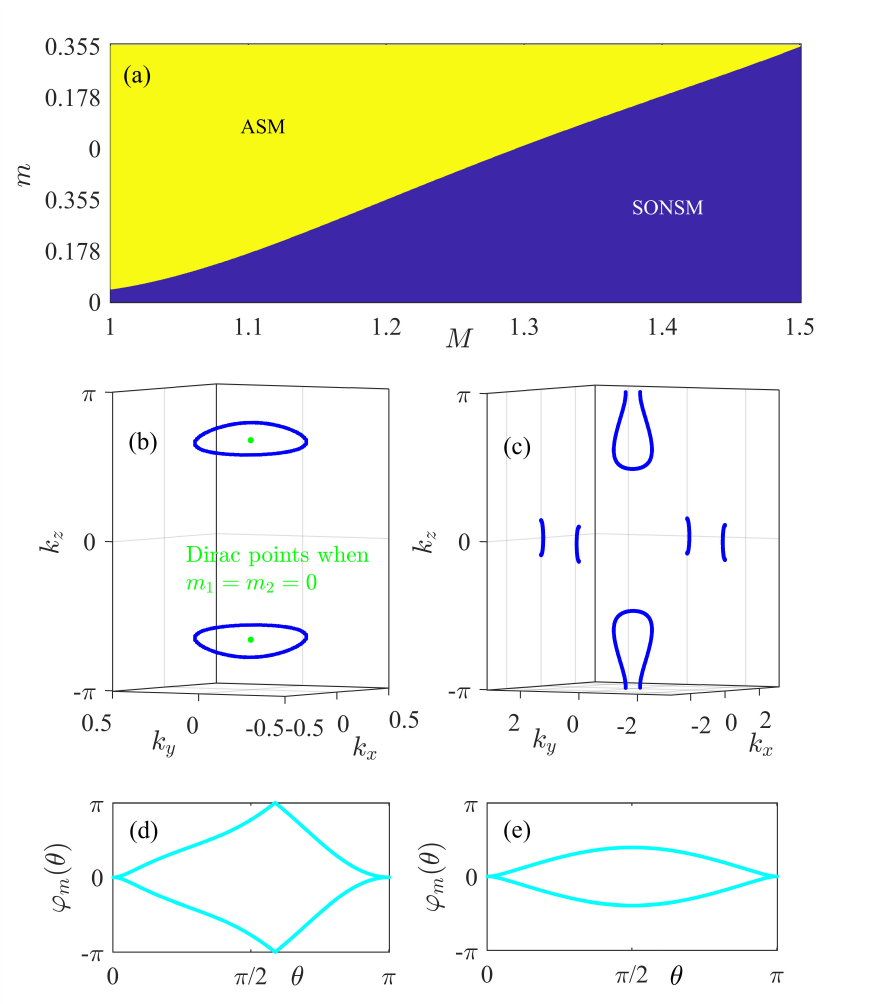}
 \caption{(a) The phase diagram in the $M-m$ plane with $m_1=m_2=m$. ASM denotes the anomalous semimetal. SONSM is {a} second-order nodal line semimetal. (b)-(c) The distributions of the nodal lines in {the} 3D Brillouin zone. The parameters are $m_1=m_2=0.2$ {for} (b) and $m_1=m_2=0.4$ {for} (c), respectively. (d)-(e) The Wilson-loop {spectra} defined on a sphere enclosing the nodal {lines} in (b) around $k_z=2.1$ and (c) around $k_z=0$, respectively. 
 The common parameter is $M=1.5$. }
\label{static1}
\end{figure}

\begin{figure}[tbp]
\centering
\includegraphics[width=1\columnwidth]{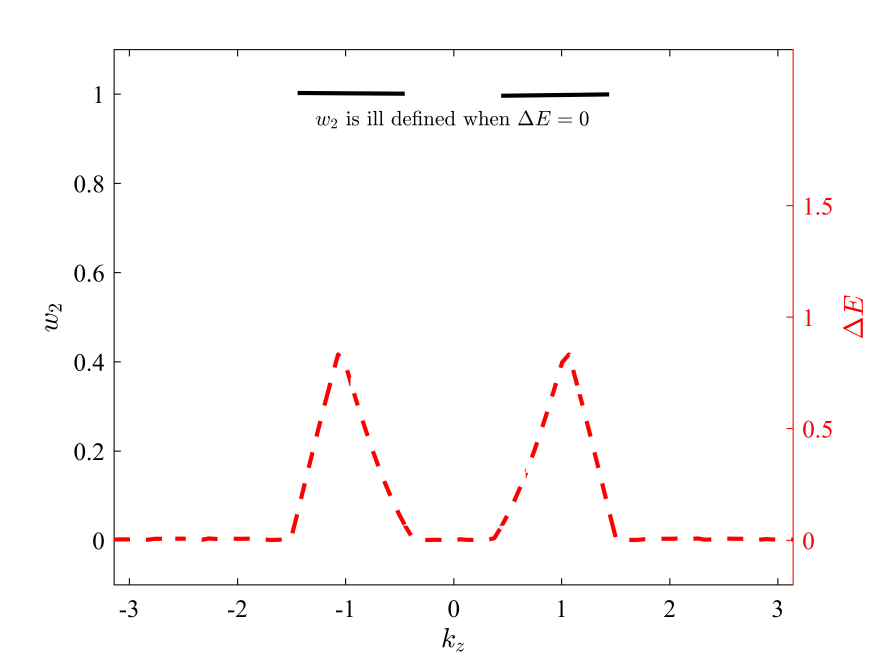}
 \caption{The $w_2$ and {corresponding energy} gap $\Delta E$ of the $k_z$-dependent 2D subsystems. The parameters are $M=1.5$ and $m_1=m_2=0.4$.}
\label{tv11}
\end{figure}

\begin{figure*}
\centering
\includegraphics[width=2.\columnwidth]{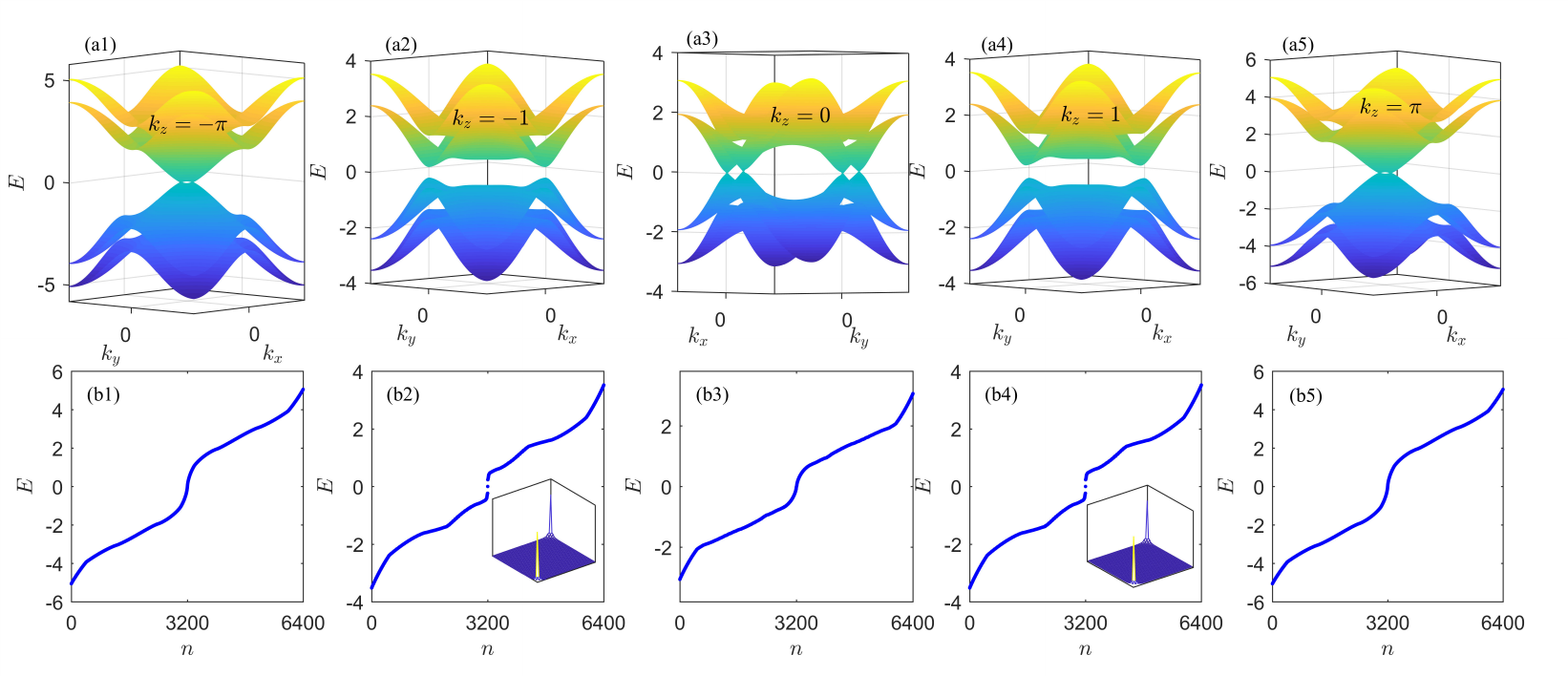}
\caption{The bulk band structures of the 2D subsystem with fixed $k_z$ in a1-a5 and the corresponding {spectra} with the open boundary condition in the $x$ and $y$ directions. The parameters are $k_z=-\pi$ in (a1-b1),  $k_z=-1$ in (a2-b2), $k_z=0$ in (a3-b3), $k_z=1$ in (a4-b4), and $k_z=\pi$ in (a5-b5). The other common parameters are $M=1.5$ and $m_1=m_2=0.4$. } \label{staticp}
\end{figure*}

\section{Static model and anomalous nodal line semimetal}
We consider a spinless system on a square {lattice with the Bloch Hamiltonian}  \cite{PhysRevLett.125.126403}
\begin{eqnarray}
\mathcal{H}(\mathbf{k})&=&\sin k_x\gamma_1+\sin k_y\gamma_2+(M-\sum_i\cos k_i)\gamma_3\nonumber\\
&+&i(m_1\gamma_1\gamma_4+m_2\gamma_2\gamma_4),
\end{eqnarray}
where $\gamma_1=\sigma_0\tau_z$, $\gamma_2=\sigma_y\tau_y$, $\gamma_3=\sigma_0\tau_x$, and $\gamma_{4,5}=\sigma_{x,z}\tau_y$. {Here} $\sigma_i$ and $\tau_i$ are two sets of Pauli matrices and $\sigma_0$ is the identity matrix.
This model with small $m_1$ and $m_2$ was investigated in Ref. \cite{PhysRevLett.125.126403}. When $m_1=m_2=0$ and $1<M<3$, $\mathcal{H}(\mathbf{k})$ describes a 3D Dirac semimetal, where there are two Dirac points residing at $k_z=\pm\arccos(M-2)$ on the $k_z$-axis. This 3D system can be sliced into a family of $k_z$-dependent topological and normal insulators. For the two 2D subsystems on either side of a chosen Dirac point, one must be a first-order topologically nontrivial insulator, while the other must be a trivial insulator. The difference between the topological numbers $w_2$ of these two subsystems is the topological charge of this Dirac point \cite{PhysRevLett.125.126403,PhysRevLett.121.106403}. It is noteworthy that the nontrivial $w_2$ of {the} Dirac point can also be calculated by considering a sphere surrounding it [see Fig. \ref{static1}(d)]. Ref. \cite{PhysRevLett.125.126403} has found that the $\mathcal{P}\mathcal{T}$-invariant perturbations $m_1$ and $m_2$ can transform the Dirac points into nodal lines [As shown in Fig. \ref{static1}(b)]. The $k_z$-dependent 2D first-order topological insulators simultaneously transform into second-order topological insulators. It is these corner modes of $k_z$-dependent phases that trace out the hinge Fermi arcs \cite{PhysRevLett.125.126403}. On the one hand, the nodal lines inherit the topological charge $w_2$ from the Dirac points. On the other hand, they carry a topological charge $w_1$,  which is defined on a small circle that transversely encircles them. $w_1$ can lead to the usual drumhead surface states. {Thus,} this nodal line semimetal hosts both usual drumhead surface states and hinge Fermi arcs that originate from the topological charges $w_1$ and $w_2$, respectively \cite{PhysRevLett.125.126403}. This phase is referred to as a second-order nodal line semimetal [see the blue area in Fig. \ref{static1}(a)]. 
 
 In addition to the already discovered second-order nodal-line semimetals \cite{PhysRevLett.125.126403}, there exists an overlooked semimetal that exhibits hinge Fermi arcs yet lacks a non-zero $w_2$ topological charge for its nodal lines [see the yellow area in Fig. \ref{static1}(a)]. As the strength of the perturbations increases, two charged nodal loops touch at the $k_z=\pi$ plane, allowing the topological charge $w_2$ to dissipate without opening a gap [see Fig. \ref{static1}(c)]. This phenomenon is entirely different from the pair creation and annihilation of Weyl or Dirac points. 
 At the same time, new nodal lines around the $k_z=0$ plane appear [see Fig. \ref{static1}(c)].  We then plot the topological invariants and {corresponding energy} gap of $k_z$-dependent 2D subsystems in Fig. \ref{tv11}. Unlike the previous case where nodal lines separate conventional and second-order topological insulators \cite{PhysRevLett.125.126403}, they now separate the same {type of} topological insulators. Fig. \ref{staticp} exemplifies the band structures for a series of subsystems with different $k_z$. When $k_z=\pm \pi$ or $0$, the corresponding 2D subsystem exhibits a Dirac semimetal [see Fig. \ref{staticp}(a1)-(b1) or (a5)-(b5)].  
 If $k_z$ is $\pm1$, the corresponding 2D subsystem is just a second-order topological insulator with $w_2=1$ [see Fig. \ref{staticp}(a2)-(b2) ]. Here, it is interesting to find {that} this anomalous nodal line semimetal hosts the hinge Fermi arc but no non-zero $w_2$ for the nodal line. The above discussion confirms that there is no correspondence between the topological charge of nodal point/line and hinge Fermi arc. 

\section{Anomalous Floquet semimetal}
To generate new semimetals, we consider a periodically driven system with the Bloch Hamiltonian given by
\begin{equation}
\mathcal{H}({\bf k},t)=\mathcal{H}_1({\bf k})+\sum_{n>0}\mathcal{H}_2({\bf k})\delta(t/T-n),\label{odr}
\end{equation}
where $\mathcal{H}_1({\bf k})=\mathcal{H}({\bf k})$, $\mathcal{H}_2=m_3\gamma_3$. $T$ is the driving period, and $n$ is an integer. {Due to the lack of energy conservation in this time-periodic system, the energy spectrum is not well-defined.} According to the Floquet theorem, the one-period evolution operator ${U}(T)=\mathbb{T}e^{-i\int_{0}^{T}\mathcal{H}(t)dt}$ defines an effective Hamiltonian $\mathcal{H}_\text{eff}\equiv {i\over T}\ln {U}(T)$ whose eigenvalues are called the quasienergies \cite{Bai_2021}. From the eigenvalue equation $U({T})\lvert u_l \rangle=e^{-i\varepsilon_{l}T}\lvert u_l \rangle$, we conclude that the quasienergy $\varepsilon_l$ is a phase factor, which is defined modulus $2\pi/T$ and takes values in the first quasienergy Brillouin zone [$-\pi/T$,$\pi/T$]. The topological phases of our periodically driven system are defined {within this} quasienergy spectrum \cite{PhysRevLett.113.236803,PhysRevB.100.085138,PhysRevB.102.041119,PhysRevResearch.2.033494,zhu2024multipletopologicaltransitionsspectral}. Unlike the static case, {topological features can emerge} at both quasienergies of $0$ and $\pi/T$. 
\begin{figure}[tbp]
\centering
\includegraphics[width=1\columnwidth]{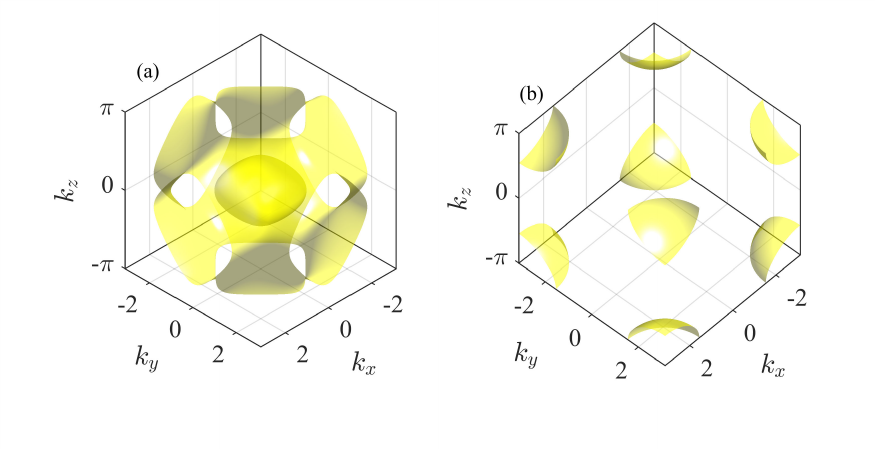}
 \caption{(a)-(b) {The distributions} of the nodal surface in \textcolor{red}{the} 3D Brillouin zone for $0$ and $\pi/T$ gap, respectively. The parameters are $M=1$, $m_1=m_2=0$, $m_3=\pi/2$, and $T=2$. The two-dimensional systems at 
 $k_z=k'_z=\pm\arccos(M+2-\frac{n\pi}{T})\approx \pm1.71$ separate \textcolor{red}{the} gapless and gapped area for $\pi$ gap. $k'_z$ is given by EQ. \eqref{wo} when $k_x=k_y=\pi$. 
}
\label{fnl}
\end{figure}
\begin{figure}[tbp]
\centering
\includegraphics[width=1\columnwidth]{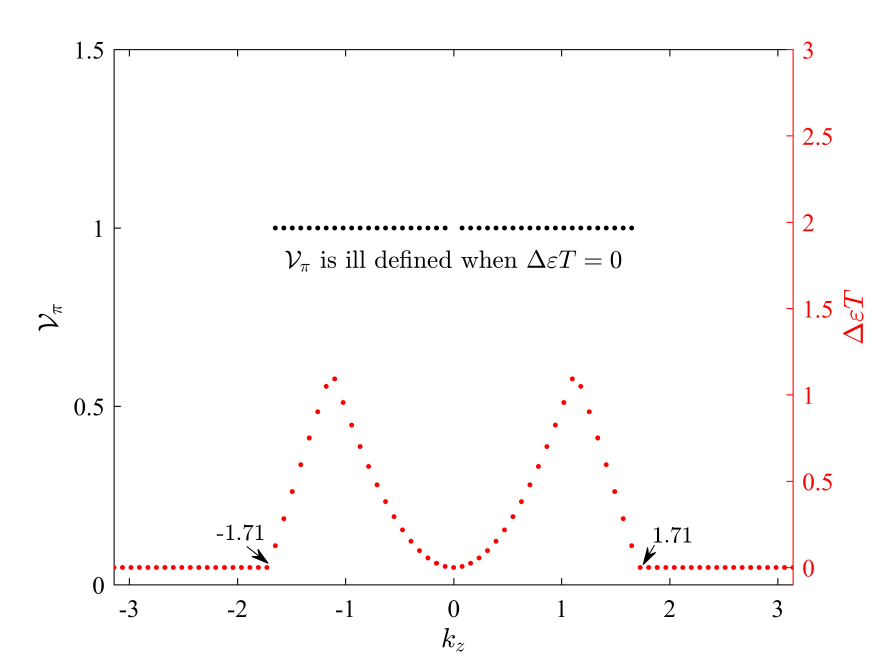}
 \caption{The $\mathcal{V}_{\pi}$ and gap $\Delta \varepsilon T$ of the $k_z$-dependent 2D subsystems. {The parameters are} $M=1$, $T=2$ and $m_1=m_2=0.$}
\label{tv22}
\end{figure}
\begin{figure*}
\centering
\includegraphics[width=2.\columnwidth]{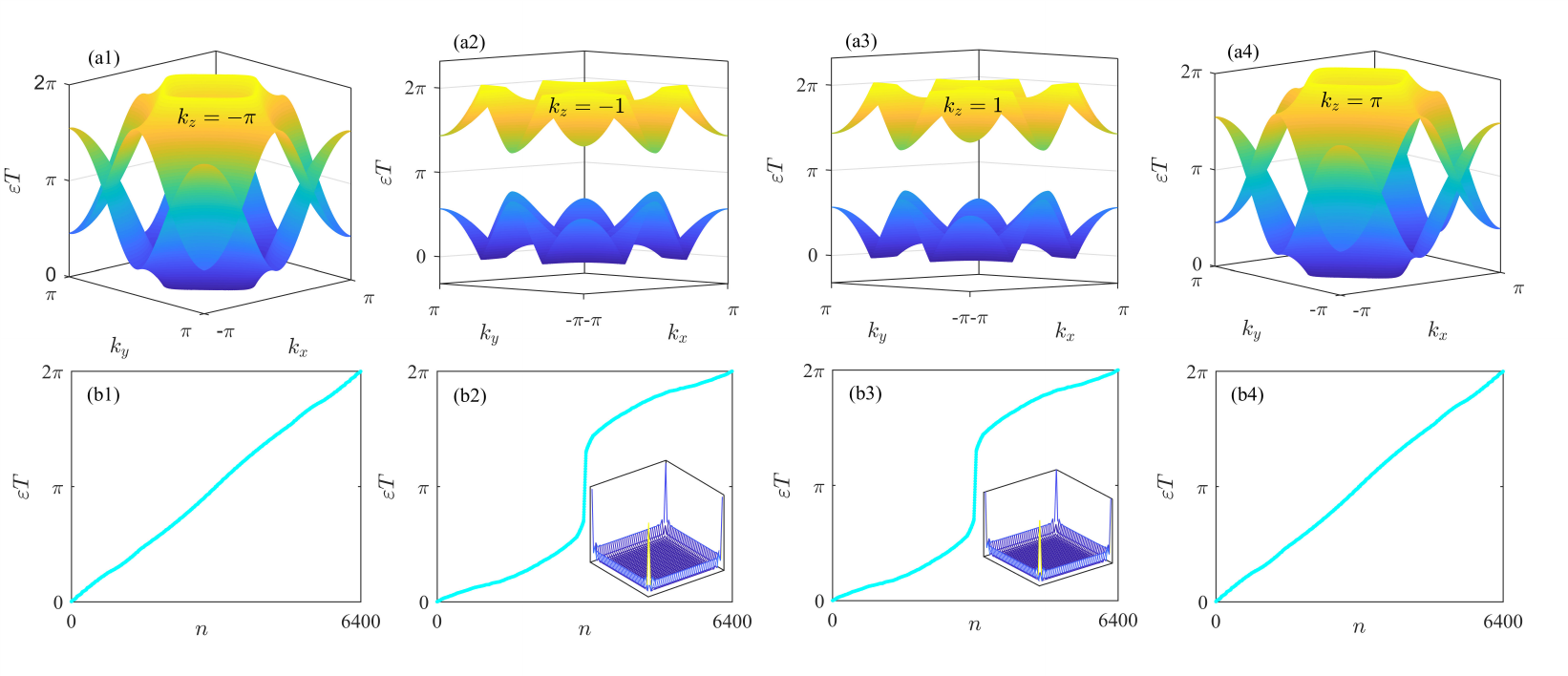}
\caption{(a1-a4) The bulk band structures of the 2D subsystem with fixed $k_z$  and (b1-b4) the corresponding {spectra} with the open boundary condition in the $x$ and $y$ directions. {The parameters are} $k_z=-\pi$ in (a1-b1),  $k_z=-1$ in (a2-b2), $k_z=1$ in (a3-b3) and $k_z=\pi$ in (a4-b4). The other common parameters are $M=1$, $m_1=m_2=0$, $m_3=\pi/2$, and $T=2$. } \label{static2}
\end{figure*}

It is straightforward to verify that both $\mathcal{H}_1({\bf k})$ and $\mathcal{H}_2({\bf k})$ exhibit space-time inversion symmetry. However, the symmetries of $\mathcal{H}_\text{eff}\equiv {i\over T}\ln {U}(T)$ may differ from those of the original static system. We propose the following scheme to resolve this problem. The space-time inversion symmetry can be inherited by $\mathcal{H}_\text{eff}$ within a symmetric time frame \cite{PhysRevB.104.205117}. This frame is obtained by shifting the starting time of the evolution backward over half of the driving period. The resulting new effective Hamiltonian is $\mathcal{H}'_\text{eff}=\frac{i}{T}\ln[e^{-i\mathcal{H}_1T/2}e^{-i\mathcal{H}_2T}e^{-i\mathcal{H}_1T/2}]$. Due to $\mathcal{K} e^{-i\mathcal{H}'_\text{eff}T}\mathcal{K}=e^{i(\mathcal{H}'_\text{eff})^{*}T}=\mathcal{K}e^{-i\mathcal{H}_1T/2}e^{-i\mathcal{H}_2T}e^{-i\mathcal{H}_1T/2}\mathcal{K}=e^{i\mathcal{H}_1T/2}e^{i\mathcal{H}_2T}e^{i\mathcal{H}_1T/2}=e^{i\mathcal{H}'_\text{eff}T}$, we can obtain that $\mathcal{H}'_\text{eff}$ is $\mathcal{P}\mathcal{T}$ invariant. 

We focus on the case where both $m_1$ and $m_2$ are zero.
 Here, it's difficult to calculate the topological charge of nodal line or surface. Therefore, new topological invariants that describe the topology of $k_z$-dependent 2D subsystems are generally needed. {This can be achieved via the following approach}. First, the static model has been studied in Ref. \cite{PhysRevLett.125.126403}, {which provides an alternative method for characterizing the topological properties of this system.} The block-diagonalized Hamiltonian of Stiefel-Whitney insulator can be obtained {through a} unitary transformation $S\mathcal{H}(k_x,k_y) S^{-1}={\rm Diag}[\mathcal{H}_{+}(k_x,k_y),\mathcal{H}_{-}(k_x,k_y)]$, where $S=\cos \frac{\pi}{4}-\sin \frac{\pi}{4}\frac{[\gamma_3,\gamma_5]}{2}$. This system can be considered as a direct sum of two Chern insulators with opposite
Chern numbers. The Chern number $\mathcal{C}_{\pm}$ of $\mathcal{H}_{\pm}(k_x,k_y)$ can then be used to count the number of gapless edge states, where 
\begin{equation}
w_2=\frac{\mathcal{C}_{+}-\mathcal{C}_{-}}{2}.
\end{equation}
 For our periodically driven system, the $k_z$-dependent 2D subsystems can also be Stiefel-Whitney insulator. However, {topological features} can occur at both of the quasienergies $0$ and $\pi/T$. Therefore, new topological invariants that describe the topologies of $k_z$-dependent 2D subsystems are generally required. To characterize this topology, {the Chern numbers} $\mathcal{C}_{\pm}$ of static case can be extended to $\mathcal{V}_{\varphi,\pm}$. The method for this extension is detailed in Ref. \cite{PhysRevB.96.195303}, which provides the topological invariants for Floquet systems. We can obtain the block-diagonalized Hamiltonian $S\mathcal{H}(\mathbf{k},t) S^{-1}={\rm Diag}[\mathcal{H}_{+}(\mathbf{k},t),\mathcal{H}_{-}(\mathbf{k},t)]$ and
\begin{eqnarray}
\mathcal{H}_{\pm}(\mathbf{k},t)&=&\sin k_x\sigma_z+\sin k_y\sigma_x\pm(M-\sum_i\cos k_i)\sigma_y\nonumber\\
          &\pm&m_3\sigma_y\sum_{n>0}\delta(t/T-n).\label{fzodr}
\end{eqnarray}
Then we can obtain $S U(t) S^{-1}={\rm Diag}[U_+,U_-]$, where $U_{\pm}=\mathbb{T}e^{-i\int_0^{t}\mathcal{H}_{\pm}(\mathbf{k},t) dt}$.
Inspired by the unique topology in periodically driven systems \cite{PhysRevLett.122.076801,PhysRevX.3.031005,PhysRevB.104.L180303}, we can construct a pair of topological {indices} $\mathcal{V}_{\varphi}$ ($\varphi$ is 0 or $\pi$) that {equals} the number of edge states at the $\varphi/T$ quasienergy. For fixed $k_z$, the evolution operator $U(k_x,k_y,k_z,t)$ can be diagonalized as diag[$U_{+}(k_x,k_y,k_z,t)$,$U_{-}(k_x,k_y,k_z,t)$]. The dynamical winding number, which characterizes the number of gapless boundary states at $\varphi/T$ ($\varphi=0$ or $\pi$) quasienergy is $\mathcal{V}_{\varphi}=\frac{\mathcal{V}_{\varphi,+}-\mathcal{V}_{\varphi,-}}{2}$ with
\begin{eqnarray}
\mathcal{V}_{\varphi,\pm}&=&\frac{1}{24\pi^2}\int_{0}^{T}dt\int dk_xdk_y\epsilon^{i_1i_2i_3}\nonumber\\&&{\rm Tr}[U^{\dag}_{\varphi,\pm}\partial_{i_1} U_{\varphi,\pm} U^{\dag}_{\varphi,\pm}\partial_{i_2} U_{\varphi,\pm} U^{\dag}_{\varphi,\pm}\partial_{i_3} U_{\varphi,\pm}]\textbf{},
\end{eqnarray}
where $U_{\varphi,\pm}=U_{\pm}({k_x},k_y,t)[U_{\pm}({k_x},k_y,T)]^{-t/T}_{\varphi}$ with
\begin{equation}
[U_{\pm}({k_x},k_y,T)]^{-t/T}_{\varphi}=\sum_{l}{\rm exp}[-\frac{t}{T}\ln_{\varphi}e^{-i\varepsilon_{\mathbf{k},l,\pm} T}]P_{l,\pm}(\mathbf{k},T).
\end{equation}
{Here, $\varepsilon_{\mathbf{k},l,\pm}T$ labels the quasienergy bands derived from $U_{\pm}(\mathbf{k},T)$ satisfying $\varepsilon_{\mathbf{k},l,\pm}T\in[\varphi,\varphi+2\varphi)$ and $P_{l,\pm}(\mathbf{k},T)$ is the projection matrix given by the eigenvector of eigenvalue $\varepsilon_{\mathbf{k},l,\pm}T$ \cite{PhysRevLett.122.076801,PhysRevX.3.031005,PhysRevB.104.L180303}.}  With these two dynamical winding numbers, we can distinguish different phases. 

Here, we choose $m_3T=\pi$, thereby yielding $\varepsilon T$  as follows
\begin{equation}
\varepsilon T=T\sqrt{\sin^2 k_x+\sin^2 k_y+(M-\sum_i \cos k_i)^2}+\pi.
\end{equation}
There are some isolated band touching points at $(k_x=0, k_y=0, k_z=\pi)$, $(k_x=0, k_y=\pi, k_z=0)$, and $(k_x=\pi, k_y=0, k_z=0)$. Expanding the Bloch Hamiltonian near the band touching points with $(q_x,q_y,q_z)$=$\mathbf{k}-\mathbf{k}_{\text{band touching point}}$ yields the low-energy effective Hamiltonian
 \begin{equation}
\mathcal{H}'_\text{eff}(\mathbf{q})=2[\pm q_x\gamma_1\pm q_y\gamma_2\pm\frac{q^2_z}{2}\gamma_3]+\pi I,\label{dineng}
 \end{equation}
 where $I$ is the identity matrix. The topological charges defined on a sphere enclosing each band touching point is zero. Therefore, we focus on the properties of the nodal surfaces. The nodal surfaces occur for $\mathbf{k}$ and the parameters satisfying 
\begin{equation}
T\sqrt{\sin^2 k_x+\sin^2 k_y+(M-\sum_i \cos k_i)^2}=n\pi,\label{wo}
\end{equation}
at the quasienergy $0$ (or $\pi/T$) if $n$ is odd integer (or even integer) 
 \cite{PhysRevB.104.205117}. Fig. \ref{fnl} displays the nodal surfaces arising from band touching points at quasienergies $0$ and $\pi/T$, respectively. Since the band touching points are distributed over $k_z\in[-\pi,\pi)$, the semimetal for zero gap is {topologically} trivial. Fig. \ref{tv22} shows the topological invariants $\mathcal{V}_{\varphi}$ 
and gap at $\pi/T$ for the $k_z$-dependent 2D subsystem. The 2D-sliced subsystems {exhibit nontrivial topological insulating behavior when} $k_z\in (-1.71, 1.71)$. Therefore, the nodal surfaces at $\pi/T$ gap are connected by topological surface states. To observe these boundary states, we plot the quasienergy spectrum and corresponding probability distributions of the edge state for the 2D subsystems in Fig. \ref{static2}. For example, {the} phase shown in Fig. \ref{static2} (b2) or (b3) for $k_z=\pm 1$ is topologically nontrivial. All crossings of the gapless edge states in these 2D subsystems contribute to the surface Fermi arc. Unlike the previous case, where nodal surfaces separate normal and topological insulators \cite{PhysRevLett.132.186601}, here they separate 2D topological insulators with {identical} $\mathcal{V}_{\pi}=1$. {Intriguingly, this} anomalous nodal semimetal hosts a surface Fermi arc but possesses no nonzero topological charge for the nodal surface.
Therefore, there is no correspondence between the topological charge of the nodal surface and Fermi arc.

\section{Discussion and Conclusion}
  In recent years, the three-dimensional Stiefel-Whitney topological phases have been realized in $\mathcal{P}\mathcal{T}$-invariant sonic and photonic crystals \cite{Xue_2023,PhysRevLett.132.197202,pan2023realhigherorderweylphotonic}. Meanwhile, periodic driving has exhibited its {versatility} in engineering exotic phases {across} various experimental platforms, including ultracold atoms \cite{RevModPhys.89.011004,PhysRevLett.116.205301,PhysRevLett.130.043201}, superconductor qubits \cite{Roushan2017}, photonics \cite{Rechtsman2013,PhysRevLett.122.173901,pan2023realhigherorderweylphotonic,PhysRevLett.133.073803}, acoustic system \cite{PhysRevLett.129.254301}. 
{We believe our theoretical predictions could be experimentally demonstrated in state-of-the-art setups} \cite{SMPnew,PhysRevLett.127.136802,PhysRevLett.119.123601,PhysRevLett.120.110603,PhysRevE.93.022209,PhysRevLett.117.220401,PhysRevLett.130.043201,PhysRevLett.102.223201,PhysRevLett.123.190603,PhysRevA.89.061603,PhysRevA.100.023622}.

In summary, we {explored} the $\mathcal{P}\mathcal{T}$-invariant semimetal and its Floquet engineering in a 3D system. {Notably,} both static nodal line and Floquet nodal surface semimetals exhibit Fermi arcs, even when the nodal lines or surfaces do not carry a $Z_2$ charge from second Stiefel-Whitney class. Such anomalous semimetals have not been discovered before. 

\section{Acknowledgments}We would like to acknowledge helpful discussions with Ken Chen. This work is supported by National Natural Science Foundation (Grants No. 12405007 and NO. 12305011), Funds for Young Scientists of Chongqing Municipal Education Commission(Grant No.KJQN20240), Natural Science Foundation of Chongqing (Grant No. CSTB2022NSCQ-MSX0316), and startup grant at Chongqing University of Posts and Telecommunications (Grant 
 No. E012A2020016, E012A2022017 and E012A2024044).

\bibliography{references}
\end{document}